\documentclass{PoS}
\include{graphics}
\title{Recent Results in AdS/QCD}

\ShortTitle{Recent Results in AdS/QCD}

\author{\speaker{Joshua Erlich}
\\
        College of William and Mary, Williamsburg, VA 23187-8795\\
        E-mail: \email{erlich@physics.wm.edu}}

\abstract{AdS/QCD is an extra-dimensional approach to modeling the light
hadronic resonances in QCD.  AdS/QCD models are generally successful at 
reproducing low-energy observables with around 10-20\% accuracy, depending
on the details of the model.  We discuss the motivation for these models,
their intrinsic limitations, and some recent results.}

\FullConference{8th Conference on Quark Confinement and the Hadron Spectrum \\
		 September 1-6 2008\\
		 Mainz, Germany}

\begin{document}

\section{What is AdS/QCD?}

AdS/QCD is an extra-dimensional approach to modeling the light
hadronic resonances in QCD, motivated by the AdS/CFT correspondence in
string theory \cite{Maldacena}.
AdS/QCD models combine several features of 
previous approaches to modeling the spectrum and interactions
of light hadrons, including:
chiral symmetry breaking,
hidden local symmetry \cite{Bando}, Large-$N$, and the Weinberg sum rules.

%

There are two complementary classes of AdS/QCD models: 
top-down models rooted in string theory,
and phenomenological bottom-up models.  The benefit of top-down models
is that both sides of the AdS/CFT duality are often well understood.  The
benefit of bottom-up models is that there is more freedom to build in
properties of QCD.

\subsection{Top-Down AdS/QCD and the AdS/CFT Corresponence}

\begin{figure}[h]
\begin{center}
\includegraphics[scale=.7,viewport=000 00 1500 170,clip]{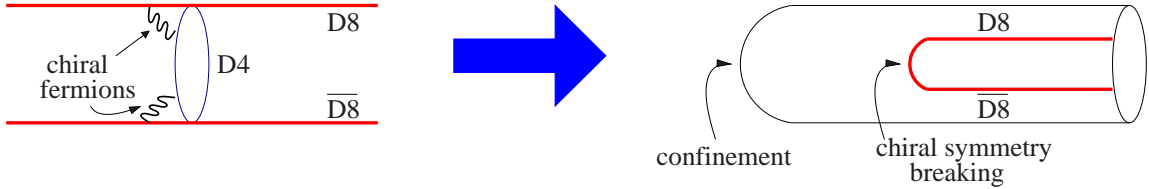}
\end{center}
\caption{Brane configuration in the Sakai-Sugimoto model.  The D4-branes
create a spacetime horizon, forcing the D8 and $\overline{{\rm D8}}$-branes
to intersect.}\label{fig:1}
\end{figure}

In the top-down approach, a
brane configuration in string theory is engineered whose low-energy spectrum
of open-string fluctuations has a known field-theoretic interpretation.  
Via the AdS/CFT correspondence, for some brane constructions 
describing large-$N$ gauge theories with large 't Hooft coupling $g^2N$,
a dual description exists in terms of supergravity on
a fixed spacetime background \cite{Maldacena}.  
The basic dictionary between the dual theories was mapped out
independently by Witten \cite{Witten} and Gubser, Klebanov and Polyakov
\cite{GKP}.
There now exists a large number of examples of field theories with supergravity
duals.  Conformal invariance and supersymmetry are not essential.  The field theory can be confining with
chiral symmetry breaking, which in those respects is similar to QCD.  
Some examples of confining theories
with known supergravity 
duals are the ${\cal N}$=1$^*$ theory of Polchinski and Strassler 
\cite{Polchinski-Strassler}, the Klebanov-Strassler cascading gauge theory
\cite{Klebanov-Strassler}, the D3-D7 system of Kruczenski {\em et al.}
\cite{Kruczenski}, 
and the D4-D8 system of Sakai and Sugimoto \cite{Sakai-Sugimoto}.

The Sakai-Sugimoto model is so far the example of the AdS/CFT correspondence
most closely related to QCD.  Ignoring the gravitational backreaction,
the system includes a stack of $N$ D4-branes wrapped on a circle on which
fermions satisfy antiperiodic boundary conditions which break supersymmetry.
From an effective 3+1 dimensional point of view, the massless spectrum includes
the SU($N$) gauge fields, but the fermions and scalars are massive.  So far
this is a model constructed 
by Witten shortly after the initial AdS/CFT conjecture \cite{Witten-circle}.
$N_f$ D8-branes and $N_f$ $\overline{{\rm D8}}$-branes
transverse to the circle on which the D4-branes wrap intersect
the D4-branes on 3+1 dimensional manifolds at definite positions along the
circle, as in Fig.~\ref{fig:1}.  The massless fluctuations of open strings
connecting the D4 and D8 or $\overline{{\rm D8}}$-branes at their intersections
describe 3+1 dimensional chiral fermions, with opposite chirality at the D8
and $\overline{{\rm D8}}$-branes.  This is the Sakai-Sugimoto model 
\cite{Sakai-Sugimoto}.

In the supergravity limit with $N_f\ll N$, the D4-branes generate a 
horizon which effectively cuts off the spacetime geometry.
The location of this horizon sets the scale of masses of those fluctuations
in the 3+1 dimensional theory, which will have the interpretation of
hadron masses.  Because of the horizon, the D8 and $\overline{{\rm D8}}$-branes
intersect, which reflects the breaking of the SU($N_f$)$\times$SU($N_f$) chiral
symmetry in the theory. For the geometry to be smooth the location of the
horizon is correlated with the size of the compact circle on which the 
D4-branes wrap \cite{Kruczenski,Witten-circle}.
The perturbative massless spectrum is that of SU($N$)
QCD with $N_f$ flavors of quarks.  However, Kaluza-Klein modes associated
with the circle direction have masses comparable to the confining scale
in this theory, so the massive spectrum of QCD-like bound states cannot be
separated (in mass) from the spectrum of non-QCD-like Kaluza-Klein modes.
The 
five-dimensional nature of the effective theory on the D4-branes becomes
apparent at the same scale as the hadron masses we are interested in.
This is an important distinction between this theory and QCD.  

If we ignore the Kaluza-Klein modes around the circle, 
the fluctuations of 4+1 dimensional 
SU$(N_f)\times$SU$(N_f)$ gauge fields on the D8
and $\overline{{\rm D8}}$-branes are identified with vector mesons,
axial-vector mesons, and pions.  The quantum numbers of the corresponding
states can be identified with symmetries of the D-brane system.  3+1 
dimensional parity,
for example, is identified with a
4+1 dimensional parity, which also exchanges the two sets of
SU($N_f$) gauge fields.  If we ignore the extra circle direction, then
the effective 3+1 dimensional action on the D8-branes describes the 
effective action for the light mesons, and easily allows for the calculation
of decay constants ($f_\pi$, $F_\rho$, {\em etc.}) and couplings ({\em e.g.}
$g_{\rho\pi\pi}$).  Most results agree relatively well with experimental
data (at around the 25\% level).  The light baryons in AdS/QCD have been
identified with solitonic configurations of the 4+1 dimensional fields,
which are closely analogous to the baryons of the Skyrme model \cite{Sakai-Sugimoto}.


\ 


\section{Bottom-Up AdS/QCD}
In the bottom-up approach, we begin with the observation that the Kaluza-Klein
modes of fields in an extra dimension might be identified with the
radial excitations of hadrons in a confining gauge theory.  As in the top-down
approach, the quantum numbers of those excitations are identified with the
transformations of the Kaluza-Klein modes under corresponding symmetries.
In the hard-wall model \cite{EKSS,DP}, we begin with a 5D SU(2)$\times$SU(2)
gauge theory.  To reproduce the discrete spectrum of hadronic excitations,
the spacetime geometry must be such that the spectrum of
Kaluza-Klein gauge fields is discrete.  
The AdS$_5$
metric can be written, \begin{equation}
ds^2=\frac{R^2}{z^2}\left(dx_\mu \eta^{\mu\nu}dx_\nu-dz^2\right), 
\label{eq:AdSmetric}\end{equation}
where $\eta_{\mu\nu}$ is th 3+1 dimensional Minkowski metric with components
diag(1,-1,-1,-1);
$R$ is the AdS curvature, which we will set to 1.  For simplicity, we take
as the geometry a slice of
AdS$_5$ between an ultraviolet cutoff scale $z=\epsilon$ and
an infrared scale $z=z_m$ related to $\Lambda_{QCD}$. 
Other choices for the geometry may better
match various aspects of QCD (such as running of the QCD coupling
and the Regge spectrum \cite{KKSS}).  The SU(2)$\times$SU(2)
gauge invariance is related to the approximate
SU(2)$\times$SU(2) chiral symmetry of the up and down quarks.  (This
may be extended to SU(3)$\times$SU(3) in order to include the
strange quark.)  In order to break the chiral symmetry
we introduce a scalar field that transforms in the bifundamental
representation of the chiral symmetry, in analogy to the operator 
$\overline{q}q$.  
If this field is arranged to have
a nonvanishing background profile, the gauge invariance is spontaneously
broken.  The 4+1 dimensional action takes the form, \begin{equation}
S=\int d^5x\sqrt{-g}\left(-\frac{1}{2g_5^2}{\rm Tr}\left(L_{MN}L^{MN}+
R_{MN}R^{MN}\right)+{\rm Tr}\left(|D_MX|^2+m_X^2|X|^2\right)\right),
 \label{eq:HardWallAction},\end{equation}
where $L_{MN}$ and $R_{MN}$ are the field strengths of the two sets of SU(2)
gauge fields and $m_X^2$ is the squared mass of the bifundamental field $X$;
contractions of indices are by the AdS$_5$ metric.

According to the AdS/CFT correspondence there is a relationship
between the scaling dimension of a 3+1 dimensional operator and the mass
of the corresponding field in the 4+1 dimensional dual theory.  If we want
we can fix the mass of the bifundamental scalar field $X$ by this dictionary,
although to do so we temporarily ignore running of the scaling dimension.
The scaling dimension of the operator $\overline{q}q$ in the
ultraviolet is three.  By the AdS/CFT 
dictionary the mass of a field $X$ with scaling dimension $\Delta$ is given
by $m_X^2=\Delta(\Delta-4)$ in units of the AdS curvature, or in our case
$
m_X^2=-3$.
The negative squared mass does not lead to instability as a result of the 
curvature of Anti-de Sitter space, since it satisfies the 
Breitenlohner-Freedman bound $m_X^2\geq-4$ in these units
\cite{Breitenlohner-Freedman}.
It is important to note that the choice $m_X^2=-3$ is made here for 
definiteness, but it is not necessary
to fix $m_X^2$ in this way.  The AdS/CFT correspondence in the classical limit
is not valid for QCD with finite $N$.  However, for definiteness
we may still choose to 
use the AdS/CFT correspondence to fix parameters in the model.

The equations of motion for the $z$-dependent scalar field background,
with the gauge fields turned off, are,
\begin{equation}
\partial_z\left(\frac{1}{z^3}\partial_z X(z)\right)+\frac{3}{z^5}X(z)=0.
 \label{eq:scalarEOM}\end{equation}
The solutions for the scalar field background are, \begin{equation}
X(z)=\left(\frac{m_q}{2}z+\frac{\sigma}{2}z^3\right), \end{equation}
where $m_q$ and $\sigma$ are arbitrary.  By the AdS/CFT correspondence,
the coefficient of the solution with divergent action has the interpretation
of the source for the corresponding operator \cite{Witten,GKP}.  
We can think of the quark mass
as a source for the operator $\overline{q}q$, so $m_q$ has the interpretation
of the quark mass (which we assume to be isospin-preserving).  Similarly,
the coefficient of the finite-action solution has the interpretation of the
expectation value of the corresponding operator, so $\sigma$ has the
interpretation of the chiral condensate $\langle \overline{q}q\rangle$.
However, since the hard-wall model is phenomenological and does not follow
from a precise AdS/CFT correspondence,
these physical identifications of the parameters $m_q$ and $\sigma$ 
are not to be taken precisely. 

If we want, we can fix $g_5$ by comparison with perturbative QCD at large
$-q^2$, although the hard-wall model is not expected to be valid at high energies.
The hard-wall prediction for the vector current two-point function 
is \cite{EKSS,DP},
\begin{equation}
i\int d^4x\,e^{iq\cdot x}\langle J_\mu^a(x)J_\nu^b(0)\rangle
=\sum\frac{F_n^2}{q^2-m_n^2}
\left(g_{\mu\nu}-\frac{q_\mu q_\nu}{m_n^2}\right)\delta^{ab},
\label{eq:JJcorr}\end{equation}
where $m_n$ is the mass of the $n^{\rm th}$ Kaluza-Klein mode,
and the decay constants $F_n$ are
determined by the kinetic mixing between the
zero-mode gauge field source for the vector current
and the $n^{{\rm th}}$ excited vector meson in the Kaluza-Klein
decomposition of the action.
The perturbative one-loop result in SU($N$) QCD with
two flavors, valid for large $-q^2$, is\begin{equation}
i\int d^4x e^{iq\cdot x}\langle J_\mu^a(x)J_\nu^b(0)\rangle
=\left(q_\mu q_\nu-g_{\mu\nu}q^2\right)\delta^{ab}\frac{N}{24\pi^2}
\log(-q^2).\label{eq:pertJJcorr}\end{equation}
Using the bulk-to-boundary propagator, the AdS/CFT correspondence gives us
a trick for performing the sum in Eq.~(\ref{eq:JJcorr}) for large $-q^2$.
For more details in the context
of AdS/QCD, I refer the reader to
Refs.~\cite{EKSS,DP}.  
The result is that the AdS/QCD prediction of the vector current two-point
function is precisely of the perturbative form of Eq.~(\ref{eq:pertJJcorr})
if we set $g_5^2=\frac{24\pi^2}{N}=8\pi^2$
when $N=3$.  
It is not necessary to fix $g_5$ by matching
to the ultraviolet, just as it was not necessary to fix the mass of the 
field X by matching to the conformal dimension of the operator $\overline{q}q$.
These choices are made for definiteness, but they
may be relaxed.
\begin{table}
\begin{center}
\begin{tabular}{|c|c|c|c|}  
\hline
Observable         & Measured                & Model \\
                   & (Central Value - MeV)                   & (MeV) \\ \hline
$m_\pi$            & 139.6        & 141 \\
$m_\rho$           & 775.8          & 832   \\ 
$m_{a_1}$          & 1230             & 1220  \\
$f_\pi$            & 92.4           & 84.0  \\
$F_\rho^{\,1/2}$   & 345               & 353   \\
$F_{a_1}^{\,1/2}$  & 433              & 440   \\
$g_{\rho\pi\pi}$   & 6.03                  & 5.29  \\ \hline
\end{tabular}
\label{tab:1}\caption{Best fit of hard-wall model to seven observables,
from Ref.~\cite{EKSS}.}
\end{center}
\end{table} 
\begin{table}
\begin{center}\begin{tabular}{|c|c|c|c|}  
\hline
Observable         & Measured      & Model \\
                   & (Central Value - MeV)         & (MeV) \\ \hline
$m_{K^*}$          & 892           & 897 \\
$m_\phi$           & 1020          & 994 \\
$m_{K_1}$          & 1272    & 1290 \\
$m_K$              & 498           & 411 \\
$f_K$              & 113           & 117 \\ 
$m_{f_2}$          & 1275          & 1236 \\
$m_{\omega_3}$     & 1667          & 1656 \\
$m_{f_4}$          & 2025          & 2058 \\
$m_\eta$           & 548           & 520 \\
$m_\eta'$          & 958           & 867 \\ 
\hline
\end{tabular}
\label{tab:2}\caption{Additional predictions of hard-wall model with three
quark flavors, from Ref.~\cite{Katz-Lat08}.}
\end{center}
\end{table} 
If we do fix $m_X^2$ and $g_5$ this way, the model has three 
remaining parameters.  A root-mean-squared fit of the remaining parameters
to the central values of
experimental and lattice data for seven observables gives $z_m=1/(346\
{\rm MeV})$, $\sigma=(308\ {\rm MeV})^3$,
$m_q=2.3\ {\rm MeV}$ \cite{EKSS}.  The hard-wall AdS/QCD predictions with these
values for the parameters are given in Table~1.  A number of
additional predictions are given in Table~\ref{tab:2}, taken from 
Emanuel Katz's talk at Lattice 2008 \cite{Katz-Lat08}.  Some of the
latter predictions required fitting an additional parameter analogous to $m_q$
in the profile of the field $X$ corresponding
to the strange quark mass.  The best fit for that
parameter takes the value $m_s=35$ MeV, which deserves some comment due to
its small value.  Recall that the parameters in the hard
wall are only related to QCD parameters to the extent that the AdS/CFT
correspondence is valid for this model.  The parameter $m_s$ is analogous
to the strange quark mass, but it is not the strange quark mass.  Furthermore,
in the fit which led to Table~\ref{tab:2}, the parameter related to the strange
quark condensate was assumed for simplicity
to be the same as that for the up and down quarks \cite{Katz-private}.  A
relation between observables and the parameters $m_q$ and $\sigma$ can
be derived which resembles the Gell-Mann--Oakes--Renner relation 
\cite{EKSS},
\begin{equation}
m_\pi^2f_\pi^2=2m_q\sigma.\end{equation}
This is a reflection of the pattern of chiral symmetry breaking, which
is built in to the AdS/QCD model.  Hence, if the strange quark condensate
violates SU(3) isospin, then the parameter $m_s$ has an even less direct
relation to the strange quark mass.
\section{Soft-Wall AdS/QCD}
The AdS/QCD models described above are not expected to be valid much above
the scale of the lightest vector resonances.  For heavy resonances,
the vector and axial-vector masses in the hard-wall model scale as
$m_n^2\sim n^2$.  However, experimental data confirm the Regge behavior
$m_n^2\sim n$.  Misha Shifman has stressed this difference between AdS/QCD
and QCD \cite{Shifman-Regge}.  It is possible to produce the Regge
spectrum by effectively modifying the AdS$_5$ geometry
\cite{KKSS}.  One way to do
this is to couple the fields in the AdS/QCD model to a background dilaton:
\begin{equation}
S=\int d^5x\sqrt{-g}e^{-\Phi(x,z)}{\cal L},\end{equation}
where the appropriate background for the dilaton is
$\Phi_0(z)\sim z^2$.  Low-energy predictions are comparable to, but not
the same as, those of the hard-wall model.

\section{Other Predictions of AdS/QCD Models}

\subsection{Form Factors}
It is straightforward to calculate form factors in AdS/QCD models,
as the AdS/CFT correspondence teaches us that the contribution of a 
tower of resonances can
be summed by use of the bulk-to-boundary propagator (see also 
Ref.~\cite{Hirn-Sanz}).
Several authors have calculated various form factors of the pion, $\rho$ and
$a_1$ mesons.
From these form factors were deduced moments
of generalizaed parton distributions, charge radii and gravitational radii.
For example, in the hard wall model with a particular choice of parameters
 \cite{Kwee-pi,Grigoryan-rho,Abidin}:
\[\langle r_\pi^2\rangle_{charge}=0.33\  {\rm fm}^2,\ \ 
\langle r_\pi^2\rangle_{grav}=0.13\ {\rm fm}^2\]
\[\langle r_\rho^2\rangle_{charge}=0.53\ {\rm fm}^2,\ \ 
\langle r_\rho^2\rangle_{grav}=0.21\ {\rm fm}^2\]
\[\langle r_{a_1}^2\rangle_{charge}=0.39\ {\rm fm}^2,\ \ 
\langle r_{a_1}^2\rangle_{grav}=0.15\ {\rm fm}^2\]
 
\begin{figure}

\includegraphics[scale=.4
]{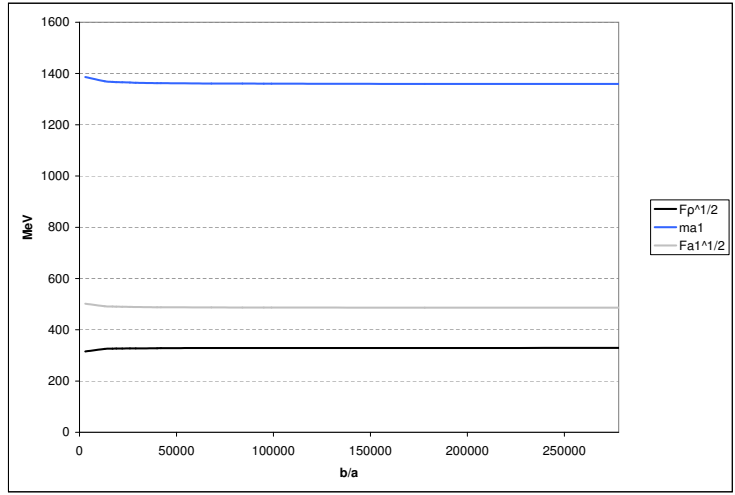}
\centering
\caption{Dependence of observables on IR boundary conditions in
hard-wall model,
holding fixed $f_\pi$, $m_\pi$, and $m_\rho$.  The boundary conditions
are of the form $a\,\partial_\nu F^{\mu \nu}(x,z_m)+b\, F^{\mu z}(x,z_m)=0$.
Plot by Christopher Westenberger \cite{Westenberger}.}
\label{fig:Westenberger}
\end{figure}

\subsection{Baryons}
If one naively truncates an AdS/QCD  model to the lightest modes, the
gauge kinetic terms of type $F_{\mu z}F^{\mu z}$ include
the Skyrme term with coefficient that depends on the 4+1 dimensional gauge
coupling and the spacetime geometry \cite{Sakai-Sugimoto}.  The light baryons have
therefore been identified as Skyrmions in AdS/QCD.  There has been
some debate as to the stability of predictions for baryons in this approach,
as in the original Skyrme model, based on the relative importance of 
higher-dimension operators that are neglected in this approach.  
A better description
in terms of 4+1 dimensional solitons, as opposed to the Skyrmions in the
effective 3+1 dimensional theory, has been discussed by several authors,
for example in Refs.~\cite{Nawa,Pomarol-skyrmions}.  An alternative description
of baryons modeled as fundamental fermions in the extra dimension has also
been considered, and has been 
seen to accurately reproduce the spectrum of excited Delta and nucleon 
resonances  (see, for example, Refs.~\cite{Brodsky-baryons,Hong-baryons}).

\subsection{Light-front wave functions}
Stan Brodsky and Guy de Teramond have noticed an intriguing relationship
between the equations of motion for Kaluza-Klein modes of fields
with general spin and scaling dimension in AdS/QCD
models, and equations describing light-front wavefunctions of
hadrons with general spin and orbital angular momentum 
({\em e.g.} Ref.~\cite{Brodsky-lightfront}).  
This is an intriguing observation, and 
relates the radial direction of Anti-de Sitter space with partonic momenta
inside the hadrons.  

\section{Universality in AdS/QCD}
Certain observables in AdS/CFT models at finite temperature have been
found to be completely independent of the details of the model.  A famous
example is the ratio of shear viscosity $\eta$ to entropy density $s$, 
which in natural
units is predicted to be \cite{Kovtun-Son-Starinets}, \begin{equation}
\frac{\eta}{s}=\frac{1}{4\pi}.\end{equation}
A more recent example is the ratio of electrical conductivity $\sigma$ to
charge susceptibility $\chi$, which depends only on
the temperature and number of spacetime dimensions \cite{Kovtun-Ritz}.
Univeral predictions provide the strongest test of the AdS/CFT correspondence
as applied to QCD.  Despite the absence of complete universality of most
AdS/QCD predictions, it is interesting that the various AdS/QCD models
make comparable predictions for low-energy observables.  
It is worthwhile to better understand which 
observables are approximately independent
of the details of the AdS/QCD model, and which details of the model are 
unimportant.  For example, upon varying the boundary condition of the
gauge fields at the infrared boundary in the hard-wall model, as long as
the remaining parameters are chosen so as to correctly reproduce a small
number of observables ($m_\pi, \ f_\pi$ and $m_\rho$) the remaining
low-energy observables of Table~1 were found to vary by only
a few percent, as in Fig.~\ref{fig:Westenberger} \cite{Westenberger}.
Perhaps the surprising success of AdS/QCD models at low energies is the
result of such universality in its predictions.

\section{Additional Applications of AdS/CFT models}
Models similar in spirit to AdS/QCD have been applied to other 
strongly-interacting dynamical systems.  Holographic technicolor models
\cite{Hirn-Sanz-holotech}
are based on
AdS/QCD, and AdS/CFT methods allow for calculation of precision
electroweak observables in these models.  These models are examples
of Higgsless models of electroweak symmetry breaking, in which 
unitarity of longitudinal W boson scattering is the result of interactions
involving the massive Kaluza-Klein excitations of the
electroweak gauge bosons.  

An exciting recent application of the AdS/CFT correspondence
is to condensed matter systems.  An important development in this direction
was the construction of a spacetime geometry whose isometries are
the same as the  nonrelativistic conformal group 
\cite{Son-nonrel}.
There are intriguing similarities of some
models to high-temperature superconductors and systems of cold atoms
\cite{Gubser-coldatoms}.

\section{Summary}
AdS/QCD models share the features of a number of earlier approaches to
modeling QCD at low energies.  AdS/QCD models generally predict low-energy
observables at the 10-20\% level, but do not fare as well at high energies.
It is not yet clear why these models work as well as they do, but some
predictions have been found to be universal as details of the
model are varied.

\acknowledgments
The author thanks Zainul Abidin, Carl Carlson, Hovhannes Grigoryan,
Emanuel Katz, Herry Kwee,
Richard Lebed, Anatoly Radyushkin and Christopher Westenberger
for use of their results in this talk.  The author's research was
supported by NSF grant PHY-0504442.  The work of C. Westenberger presented
here was supported by the NSF through
the Research Experiences for Undergraduates program at the College of
William \& Mary.

\end{document}